\documentclass[aps,prl,twocolumn,nofootinbib,superscriptaddress]{revtex4-2}

\pdfoutput=1

\usepackage[utf8]{inputenc} 
\usepackage{amsmath}
\usepackage{amssymb}
\usepackage{xcolor}
\usepackage{graphicx}
\usepackage{dcolumn}
\usepackage{bm}
\usepackage{microtype}
\usepackage{threeparttable}
\usepackage{mathtools}
\usepackage[colorlinks=true]{hyperref}
\usepackage{physics}
\usepackage{comment}
\usepackage{soul} 
\usepackage{bbold} 
\usepackage{ragged2e}

\newcommand{\be}{\begin{eqnarray}}
\newcommand{\ee}{\end{eqnarray}}

\begin{document}

\title{Krylov spread complexity as holographic complexity beyond JT gravity}

\author{Michal P. Heller}
\email{michal.p.heller@ugent.be}
\affiliation{Department of Physics and Astronomy, Ghent University, 9000 Ghent, Belgium}

\author{Jacopo Papalini}
\email{jacopo.papalini@ugent.be}
\affiliation{Department of Physics and Astronomy, Ghent University, 9000 Ghent, Belgium}

\author{Tim Schuhmann}
\email{tim.schuhmann@ugent.be}
\affiliation{Department of Physics and Astronomy, Ghent University, 9000 Ghent, Belgium}

\begin{abstract}
\noindent One of the important open problems in quantum black hole physics is a dual interpretation of holographic complexity proposals. To date the only quantitative match is the equality between the Krylov spread complexity in triple-scaled SYK and the complexity = volume proposal in classical JT gravity. Our work utilizes the recent connection between double-scaled SYK and sine dilaton gravity to show that the quantitative relation between Krylov spread complexity and complexity = volume extends to finite temperatures and to full quantum regime on the gravity side at disk level. Furthermore, we isolate the first quantum correction to the complexity = volume proposal and propose to view it as a complexity of quantum fields in the bulk. The key lesson from our work is that gravity demands to assign complexity also to Euclidean state preparation.
\end{abstract}

\maketitle

\noindent \textbf{\emph{Introduction.--}} Understanding black hole interiors has been an important theme of research in quantum gravity ever since the foundational papers on black holes appeared. A major modern incarnation of this research front has been studies of holographic complexity over the course of the past decade~\cite{Susskind:2018pmk,Chapman:2021jbh}. This research direction stems from a heuristic yet convincing observation in~\cite{Susskind:2014rva} that connects perpetual growth of black hole interiors in spatial volume with the expected behavior (proven in random quantum circuits~\cite{Haferkamp:2021uxo}) of quantum circuit complexity in chaotic systems. In the realm of holography this property of black hole interiors is now formalized within the complexity = anything paradigm~\cite{Belin:2021bga,Belin:2022xmt}, which provides an infinite set of boundary-anchored bulk geometric objects which all grow linearly with the boundary time at late enough times. Arguably the simplest complexity = anything proposal is the one that was considered first: the complexity = volume proposal dealing with the spatial volume of extremal time slices of the bulk~\cite{Stanford:2014jda}.

The key open question in the studies of holographic complexity is that of its interpretation in the microscopic description on the boundary. To date, despite a decade of research on the topic there has been only one fully quantitative match discovered \cite{footnote1}.
In 2023 the authors of~\cite{Rabinovici:2023yex} (see also 2022~\cite{Lin:2022rbf}) showed that complexity = volume in the simplest holographic gravity theory (classical JT gravity) is given by the Krylov spread complexity~\cite{Balasubramanian:2022tpr} in the dual description in terms of a particular (triple-scaling) limit of the SYK model at infinite temperature. While this may come as a surprise, since Krylov complexity is a priori a distinct quantity from quantum circuit complexity originally proposed in this context, both quantities realize Susskind's expectation about linear growth behavior in chaotic quantum systems~\cite{Balasubramanian:2019wgd,Haferkamp:2021uxo,Balasubramanian:2022tpr}. 

The so far isolated nature of~\cite{Rabinovici:2023yex} raises the crucial question: \textit{does the relation between holographic complexity and microscopic Krylov spread complexity extend to other gravity setups?} The aim of our letter is to present that the answer is affirmative within a freshly discovered duality between two-dimensional sine dilaton gravity and a double-scaled SYK model \cite{Blommaert:2024ydx,Blommaert:2023opb,Blommaert:2024whf,Blommaert:2025avl}. From the point of view of the parameter space of the double-scaled SYK model, the result of~\cite{Rabinovici:2023yex} lies at a single point and our letter extends it to a two-dimensional plane spanned by the interaction strength (nonlocality of the model), in which case we make use of the quantum generalization of the complexity = volume proposal put forward in~\cite{Iliesiu:2021ari}, and inverse temperature $\beta$. In the outlook we speculate on the implications of our study for the broader field of holographic complexity.\\

\noindent \textbf{\emph{Setup.--}}
The SYK model~\cite{Sachdev:1992fk,kitaevvideo} describes a system of N Majorana fermions, with dynamics governed by a p-body interaction of the form:
\begin{equation}
\label{eq:H_syk}
H_{\mathrm{SYK}} = i^{p/2} \sum_{1 \leq i_1 < \cdots < i_p \leq N} J_{i_1 \cdots i_p} \psi_{i_1} \cdots \psi_{i_p} \,,
\end{equation}
where the couplings $J_{i_1 \cdots i_p}$ are typically assumed to be Gaussian random variables.
We will be focusing on the double-scaled SYK model (DSSYK), where both $N$ and~$p$ are sent to infinity, while keeping the ratio $\abs{\log q} = \frac{p^2}{N}$
finite. A striking feature of this regime is that all amplitudes in DSSYK are computable in an exact way, by evaluating the so called chord diagrams, an intermediate combinatorial tool to solve the system \cite{Berkooz:2018jqr,Berkooz:2018qkz}. Specifically, this is done by introducing an auxiliary quantum mechanical system governed by an effective Hamiltonian, the transfer matrix $\hat{T}$, which acts on a Hilbert space spanned by chord states $\ket{n}$. See for instance \cite{Berkooz:2024lgq} for a recent comprehensive review.

Our gravitational setup is sine dilaton gravity, a recently proposed gravity dual to double-scaled SYK \cite{Blommaert:2024ydx,Blommaert:2023opb}. The gravitational path integral of sine dilaton gravity takes the form
\small
\begin{equation}\label{b+b}
    \int \mathcal{D} g \mathcal{D}\Phi\,\exp\bigg( \frac{1}{2}\int \mathrm{d}^2 x \sqrt{g}\bigg(\Phi R+\frac{\sin(2\left|\log q\right|\Phi)}{\left|\log q\right|}\bigg)\bigg)\,,
\end{equation}
\normalsize
where we suppressed the boundary terms. In~\eqref{b+b} the  potential for dilaton $\Phi$ exhibits a sine profile and thus corresponds to a deformation of the linear dilaton potential of JT gravity, see e.g.~\cite{Mertens:2022irh} for a review of the latter. The main object of interest for us is the extremal volume of the complexity = volume proposal in this gravity theory, which in two dimensions is just the geodesic length. Specifically, to prove that DSSYK Krylov spread complexity is dual to length in this gravity theory, we will be interested in the gravitational two-point function in sine dilaton gravity, corresponding \textcolor{black}{to} the boundary to boundary propagator of a non-minimally coupled scalar field in the bulk \cite{Blommaert:2024ydx}. The effective geometry this matter probe will experience is given by
\small
\begin{equation}\label{Ads}
    \mathrm{d} s_{\mathrm{eff}}^{2}=\mathrm{d} s_{\mathrm{sine}}^{2} e^{-2i \left| \log q \right| \Phi}=-\left(\rho^2-\sin(\theta)^2 \right)\mathrm{d} t_{S}^2+\frac{\mathrm{d}\rho^2}{\rho^2-\sin(\theta)^2},
\end{equation}
\normalsize
which represents the Schwarzschild patch of an AdS$_2$ black hole with Hawking temperature $\beta_{\mathrm{BH}}=2\pi/\sin(\theta)$, which reduces to the JT black hole when $\theta \ll 1$. \textcolor{black}{For the remainder of this letter, we will use the parameter $\theta$ to conveniently parametrize finite temperatures.} The length in the Kruskal extension of the geometry~\eqref{Ads}, after holographic renormalization, is given by
\begin{equation}\label{length}
    L=2 \log \left(\cosh \left(t\sin (\theta )/2 \right)\right)-2\log (\sin (\theta ))
\end{equation}
in terms of the two-sided boundary Lorentzian time $t$.
The expression \eqref{length} represents the semiclassical limit of the full quantum expectation value of the length of the thermofield double state~\cite{Maldacena:2001kr} in sine dilaton gravity as a function of $t$, which we will compute in \eqref{eq:intermediate_step} and match with Krylov spread complexity in DSSYK.
By introducing the canonical conjugate $P$ of the length $L$ and performing canonical quantization, the gravitational Hamiltonian $\hat{H}_\text{grav}$ of sine dilaton gravity reads \cite{Blommaert:2024ydx,Blommaert:2024whf} 
\small
\begin{equation}\label{grav2}   \hat{H}_\text{grav}=\frac{1}{\sqrt{2 \left|\log q\right|}\sqrt{1-q^2}}\left[-\cos(\hat{P})+\frac{1}{2}e^{\mathrm{i} \hat{P}}e^{-\hat{L}}\right],
\end{equation}
\normalsize
which precisely matches the DSSYK transfer matrix $\hat{T}$ \cite{Berkooz:2018jqr,Berkooz:2018qkz,Lin:2022rbf}, making the duality manifest. Moreover, as discussed in detail in \cite{Blommaert:2024whf}, the invariance of the Hamiltonian \eqref{grav2} under periodic shifts $P\rightarrow P+2\pi$ of the momentum instructs us to consider this as a redundancy and to treat $P$ as a compact variable. Upon doing so, the theory is projected onto discretized lenghts \textcolor{black}{and the Euclidean time periodicity $\beta_{BH}$ of the black hole background \eqref{Ads} reduces to the true microscopic temperature of DSSYK $\beta=(2\pi-4\theta)/\sin(\theta)$} \cite{footnote2}.
We point out how the discretization of chord number is obvious from the point of view of DSSYK \cite{Berkooz:2018jqr,Berkooz:2018qkz,Lin:2022rbf}, but the quantization of the length is far from obvious from the gravity side. Upon projection, the physical Hilbert space $\ket{L}$ can then be identified with the chord Hilbert space of DSSYK, spanned by the chord number states $\ket{n}$ \cite{Berkooz:2018jqr,Berkooz:2018qkz,Lin:2022rbf}, via the holographic dictionary
\begin{equation}\label{holo}
L=2 \left| \log q\right|n.
\end{equation}
More details on the presented duality can be found in the supplemental material \cite{SM} in Sec. SM1.

In sine dilaton gravity, the two-point function of a massive probe is captured by the insertion of the bilocal operator $e^{-\Delta \hat{L}}$ of conformal weight $\Delta$ \cite{Blommaert:2024ydx}. 
The structure of this expectation value takes a similar form to the JT gravity one \cite{Blommaert:2018oro,Iliesiu:2019xuh} and reads
\small
\begin{equation}\label{2exact}
\hspace{-8 pt}\langle e^{-\Delta \hat{L}} \rangle=Z_\beta^{-1}\bra{L=0}e^{-\tau \hat{H}_\text{grav}}\,e^{-\Delta \hat{L}}\,e^{-(\beta-\tau)\hat{H}_\text{grav}} \ket{L=0},
\end{equation}
\normalsize
where the disk has been divided in two parts of length $\tau$ and $\beta-\tau$ corresponding to the Euclidean boundary times to which the semiclassical geodesic is anchored. An explicit evaluation of \eqref{2exact} is provided in the supplemental material \cite{SM} in Sec. SM2, \textcolor{black}{ including definitions of the appearing special functions}. Analytically continuing $\tau=\beta/2+i \,t$ in \eqref{2exact}, we finally obtain the Lorentzian two-point function in sine dilaton gravity. Consistently, the latter is given by the expectation value of the bilocal operator in the state
\begin{equation}
\ket{\psi(t)}_\beta=e^{-i t \hat{H}_\text{grav}}\ket{\psi_\beta} \,,
\end{equation}
which correponds to the Lorentzian time evolution of the Hartle-Hawking state $\ket{\psi_\beta}=Z_\beta^{-1/2}e^{-\beta \hat{H}_\text{grav}/2} \ket{L=0}$, obtained by the Euclidean half-disk preparation amplitude.\\

\noindent \textbf{\emph{Deriving Krylov complexity from length.--}}
Krylov spread complexity \cite{Balasubramanian:2022tpr} is defined as the expectation value of the position of a state
$\vert \psi(t)\rangle$ spreading over the Krylov basis $\vert K_n \rangle$
\begin{align}
\label{eq:Krylov_complexity}
    C_K(t)=\sum_{n=0}^\infty n \,\vert \langle K_n \vert \psi(t)\rangle\vert^2\,,\quad \vert \psi(t)\rangle = e^{-i H t}\vert R \rangle\,.
\end{align}
The state $\vert \psi(t)\rangle$ follows Hamiltonian time evolution from the reference state $\vert R \rangle$ and the Krylov basis $\vert K_n \rangle$ is the unique basis that arises from Gram-Schmidt orthonormalization of the set of states constructed via repeated application of the Hamiltonian to the reference state $\{H^n\vert R \rangle\}$, commonly referred to as the Lanczos algorithm.

We begin the derivation of the core statement of this letter on the gravity side and study the expectation value of the length in sine dilaton gravity. Specifically, we extract the length from the two point function \eqref{2exact}, where $\Delta$ is the scaling dimension of the bilocal operator, via
\begin{align}\label{relation}
    \langle \hat{L}\rangle=\left[-\partial_\Delta \langle e^{-\Delta \hat{L}}\rangle\right]_{\Delta=0}.
\end{align}
Note that this starting point coincides with the way the authors of \cite{Iliesiu:2021ari} lay out their study of the black hole interior at very late times in JT gravity. The key observation we make is that, by leveraging the relation \eqref{relation} to derive $\langle \hat{L}\rangle$
from \eqref{2exact} 
supplied with the analytic continuation $\beta/2+it$, one obtains
\small
\begin{align}
\label{eq:intermediate_step}
     &\hspace{-4 pt}\langle \hat{L}\rangle=\frac{\left|\log q^2\right|}{Z_\beta}\int_0^\pi \mathrm{d}\theta_1\,\rho(\theta_1)\int_0^\pi \mathrm{d}\theta_2\,\rho(\theta_2)e^{-\left(\frac{\beta}{2}+i t\right) E(\theta_1)}\nonumber\\
&\hspace{-4 pt}e^{-\left(\frac{\beta}{2}-i t\right)E(\theta_2)}\sum_{n=0}^{\infty} \frac{n}{(q^2; q^2)_n} H_n \left( \cos \theta_1 | q^2 \right) H_n \left( \cos \theta_2 | q^2 \right)
\end{align}
\normalsize
for the expectation value of the length. Importantly, a sum over $n$ appears. This sum originates from the matrix element $\bra{\theta_1} e^{-\Delta \hat{L}} \ket{\theta_2}$, provided explicitely in eq. (S12) in the supplemental material \cite{SM}, and allows us to recast this result into the form of a Krylov spread complexity \eqref{eq:Krylov_complexity} as we will show now. On the way, crucially, we identify the state $|\psi(t)\rangle$ that this Krylov spread complexity is associated to.

As discussed in the setup and further detailed in the supplemental material \cite{SM} in Sec. SM1, on the basis of the previously established duality with its holographic dictionary \eqref{holo} which relates length in sine-dilaton gravity and chord number in DSSYK, we can now identify the components appearing in \eqref{eq:intermediate_step} as eigenvalues $E(\theta)$ and chord basis eigenfunctions $\langle n|\theta\rangle $ (explicitely provided in the supplemental material \cite{SM} in eq. (S6) \cite{footnote3} and (S9), respectively) of the transfer matrix $\hat{T}$, the effective Hamiltonian of the chord Hilbert space of DSSYK \cite{Berkooz:2018jqr,Berkooz:2018qkz,Lin:2022rbf} which governs the transition from chord states $\ket{n}$ to $\ket{n\pm1}$.
The second ingredient that facilitates recasting the sine dilaton bulk length \eqref{eq:intermediate_step} into a Krylov spread complexity is the crucial insight in the works \cite{Lin:2022rbf,Rabinovici:2023yex}
that the chord basis $\ket{n}$ coincides with the Krylov basis $\ket{K_n}$ for transfer matrix evolution starting from the zero chord state as reference state. The conceptual reason for this is that, when constructing the Krylov basis, each application of $\hat{T}$ to $\ket{0}$ adds a chord such that $\ket{K_n}=\ket{n}$, where importantly the built-in orthonormalization in the Lanczos algorithm cancels out the chord annihilation contribution in $\hat{T}$ at each step \cite{footnote3.2}. Analytic knowledge of the Krylov basis in this case is essential because it allows us to precisely
identify the length \eqref{eq:intermediate_step}, evolving in Lorentzian time $t$ at nonzero DSSYK inverse temperature $\beta$, with the Krylov spread complexity $C_K(t)_\beta$
\small
\begin{align}
\label{krylov_result}
    &\langle \hat{L}\rangle= \frac{\left|\log q^2\right|}{Z_\beta}\sum_{n=0}^\infty n\left|\int_0^\pi \mathrm{d}\theta\,\rho(\theta) e^{-\left(\frac{\beta}{2}+i t\right) E(\theta)} \langle n | \theta \rangle \langle \theta | 0 \rangle \right|^2\nonumber\\
    &=\left|\log q^2\right|\sum_{n=0}^{\infty} n\left| \langle n |\frac{e^{-iT(t - i \beta/2)}}{\sqrt{Z_\beta}}  | 0 \rangle \right|^2 = \left|\log q^2\right| C_K(t)_\beta
\end{align}
\normalsize
of the (normalized) state $\ket{\psi(t)}_\beta=Z_\beta^{-1/2}e^{-iT(t - i \beta/2)} | 0 \rangle$ that arises by mixed Lorentzian plus Euclidean transfer matrix evolution from the zero chord state $\ket{0}$ as reference state. The matching
\begin{equation}
\label{eq:result}
    2\left|\log q\right|C_K(t)_\beta=\langle \hat{L}\rangle=\left[-\partial_\Delta \langle e^{-\Delta \hat{L}}\rangle\right]_{\Delta=0}
\end{equation}
is one of the main results of this letter. Note that this is a match on the quantum level that is valid for all allowed values of $q\in(0,1)$ in DSSYK, as well as for all values of~$\beta$. We refer to Sec. SM3 in the supplemental material \cite{SM} for an operator perspective on this result\textcolor{black}{: For transfer  matrix evolution starting from the zero chord state, we can identify the associated Krylov operator with the length operator of sine dilaton gravity, making complexity equals volume manifest as a statement about linear operators. In Sec. SM4, we provide} an interpretation of the identified evolution in the chord Hilbert space on the level of the double-scaled SYK model based on \cite{Rabinovici:2023yex}.\\

\begin{figure}
    \centering
    \includegraphics[width=\linewidth]{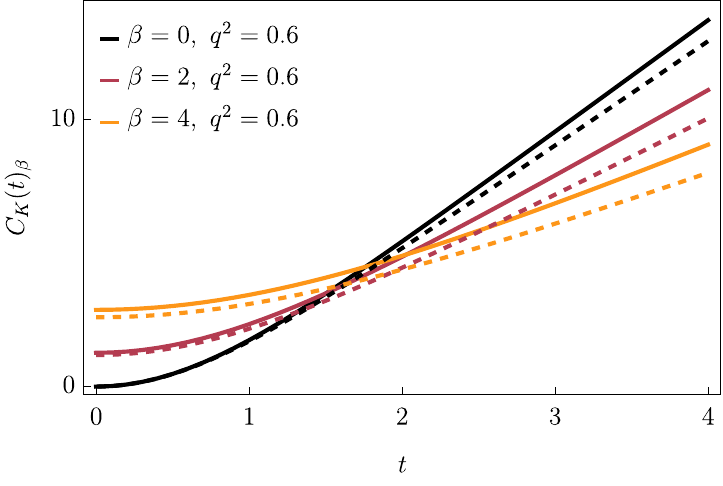}
    \caption{Numerical values for the finite temperature Krylov spread complexity $C_K(t)_\beta$ at various temperatures $\beta$ and $q^2=0.6$ (solid). It is visible that for larger $\beta$, the initial onset at $t=0$ that accounts for higher complexity of the Euclidean state preparation is larger and the complexity growth slows down. The complexities are contrasted with the expectation value of the classical length in sine dilaton gravity at temperature $\beta$ (dashed), normalized by $2\left|\log q \right|$.}
    \label{fig:finite_temp_C}
\end{figure}

\noindent \textbf{\emph{Two limits.--}} 
Although our identification \eqref{eq:result} holds at the quantum level, it is important to check that the correct semiclassical limit, as $q\to1$ limit in the setup of this work \cite{footnote4},
is recovered. Indeed, as one can observe in Fig.~\ref{fig:finite_temp_C}, the numerical evaluation of~\eqref{eq:intermediate_step} is well interpolated by the semiclassical effective length in sine dilaton gravity given in \eqref{length}. Details on the numerical evaluation can be found in the supplemental material \cite{SM} in Sec. SM5 and we defer the reader to the section about quantum corrections in the latter part of this letter for a discussion of the deviations from \eqref{length} in the quantum regime. 

The semiclassical limit was also checked for the exact two-point function in \cite{Goel:2023svz}, leading, by extrapolation through \eqref{relation}, to the same result \eqref{length}. 
We can verify that the correct JT limit is recovered by additionally considering small values of~$\theta$, where the spectral edge of the DSSYK spectrum is probed. In this regime, the semiclassical length simplifies to 
\begin{equation}\label{length_JT}
    L_{\mathrm{JT}}=2 \log \left(\cosh \left(t \theta /2 \right)\right)-2\log (\theta)
\end{equation}
which corresponds to the length of the ER bridge in the JT black hole background \cite{Harlow:2018tqv}, at temperature $\beta_{\mathrm{BH}}=2\pi/\theta$ \cite{footnote5}.
On the other hand, we can also investigate the regime where $\beta\rightarrow 0$ which, according to $\beta=\beta_{\mathrm{DSSYK}}=\frac{2\pi-4\theta}{\sin(\theta)}$, corresponds to $\theta=\pi/2$, the midpoint of the DSSYK spectral support. In this regime, the length becomes 
\begin{equation}\label{length_zerobeta}
    L_{\beta=0}=2 \log \left(\cosh \left(t /2 \right)\right)\,.
\end{equation}
In \cite{Rabinovici:2023yex}, DSSYK Krylov spread complexity for $\beta=0$ was connected to a JT classical length by performing a particular combination of limits. We clarify that taking purely the $\beta\to0$ limit does not lead to JT gravity. At the level of sine dilaton gravity, the $\beta\to0$ regime is different from the standard JT limit, as it corresponds to an expansion of the dilaton potential around $\Phi=\pi/2$. In this case, as shown in \cite{Blommaert:2024whf}, the gravity theory reduces to a regularization of flat-space JT gravity, reproducing the results of \cite{Almheiri:2024xtw} associated to the high-temperature limit of DSSYK. This clarification does not contradict the results of \cite{Rabinovici:2023yex} because their combination of limits is different from the pure $\beta\to0$ limit that we take. \\

\noindent \textbf{\emph{Lessons for holographic complexity.--}} 
There are two main lessons to be drawn from the result \eqref{eq:result} with respect to its validity at finite temperatures, a regime of precise matching of bulk and boundary complexity that is established in this letter for the first time. 

The first lesson is that \textit{gravity demands to assign complexity to both Lorentzian evolution as well as Euclidean state preparation} in order to match Krylov complexity to the Lorentzian evolution of the volume $\langle\hat{L}\rangle$ at nonzero inverse temperature $\beta$. Importantly, this notion of complexity is different from starting at a reference state proportional to $e^{-\beta T/2}\ket{0}$ and assigning complexity only to its Lorentzian evolution. 
This is because the different choice of reference state leads to a different Krylov basis and Krylov complexity:
\begin{align*}
    C_K\text{ for } e^{-iT(t-i\beta/2)} \underbracket{| 0 \rangle}_{\ket{R}} \neq C_K\text{ for } e^{-iTt}\underbracket{e^{-\beta T/2} | 0 \rangle}_{\ket{R}}\,.
\end{align*}
Our approach in this letter is well motivated by the fact that it is known from the gravitational point of view that the length at finite temperature arises via the analytic continuation $\beta/2+it$ in \eqref{eq:intermediate_step} and not by a structurally different calculation. 

The second lesson concerns the dependence of the Krylov complexity $C_K(t)_\beta$ on the temperature. We observe two important features, both clearly visible in Fig. \ref{fig:finite_temp_C}. The complexity (solid line) acquires a positive onset that increases with increasing $\beta$. This is the complexity associated to the Euclidean preparation of the Hartle-Hawking state $e^{-\beta T/2} | 0 \rangle$. The other distinct feature is that finite temperature slows down the complexity growth. This agrees with the observations in \cite{Xu:2024gfm} and, as discussed there, is consistent with what has been confirmed for other instances of the SYK model. When contrasted with the analytic result for the classical length (dashed line) given in eq. \eqref{length}, we observe correlation but not a precise match for a generic choice of~$q$ and~$\beta$. We will now show in the following section that this is expected and can be resolved by taking into account quantum corrections to the classical length.\\

\noindent \textbf{\emph{The role of quantum corrections to the length.--}}
In~\cite{Rabinovici:2023yex}, agreement between Krylov spread complexity in triple-scaled SYK and length in classical JT gravity was reported.
\begin{figure}
    \centering
    \includegraphics[width=\linewidth]{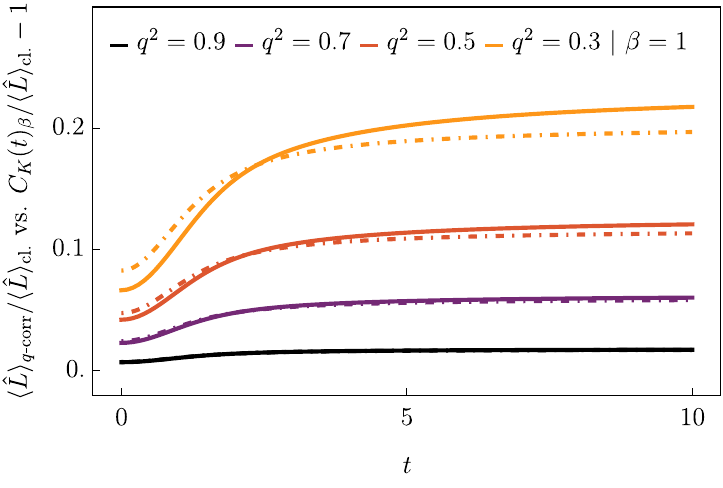}
    \caption{Numerically evaluated Krylov complexity $C_K(t)_\beta$ (solid), normalized by the classical length $\langle \hat{L}\rangle_{\text{cl.}}$ (and a factor of $2\left|\log q \right|$ that is omitted in the label) for $\beta=1$ and various $q$ (color coding) vs. the leading order quantum correction $\langle \hat{L}\rangle_{q\text{-corr.}}$ (dot-dashed), also normalized by $\langle \hat{L}\rangle_{\text{cl.}}$, for the same values of $q$ and $\beta$. For $q$ close to one, the curves fall on top of each other, signaling perfect agreement; For  $q$ further away from $q=1$, the deviation systematically increases because of higher order corrections that were not taken into account.}
    \label{fig:correction}
\end{figure}
This identification requires the limit $q\to1$ as part of the triple-scaling procedure. Let us highlight that this is very different from what is report in this letter. Here, we explicitly derive the dual Krylov spread complexity directly from the length in sine dilaton gravity, valid for all values of $q$. The strength of our result lies in the fact that we leverage a full gravitational dual to DSSYK, extending beyond the triple-scaling limit, and uncover a fully quantum bulk manifestation of complexity within the dual gravity theory. Let us focus on the region of $q$ close to but not exactly one to exemplify this statement. 

In this sector, the full quantum expression of the length in sine dilaton gravity is well approximated by the classical answer $\langle \hat{L}\rangle_{\text{cl.}}$, given in eq.\,\eqref{length}, modified by the first quantum correction in $\left|\log q\right|$. The analytical expression for this correction has been very recently derived by one of us and his collaborators in \cite{Bossi:2024ffa} and reads
\small
\begin{align}
\label{eq:q-correction_length}
    &\langle \hat{L}\rangle_{q\text{-corr.}}=
    -\frac{2\left|\log q\right|}{4 ((\pi -2 \theta ) \cot (\theta )+2)}\nonumber\\
    &  \bigg[(-t^2 \sin ^2(\theta ) \left(\cot ^2(\theta )+\tanh ^2\left(\frac{1}{2} t \sin (\theta )\right)\right)\nonumber\\
    &+\frac{4 \left(t \sin (\theta ) \tanh \left(\frac{1}{2} t \sin (\theta )\right)-2\right)}{(\pi -2 \theta ) \cot (\theta )+2}\nonumber\\
    &-((\pi -2 \theta ) \cot (\theta )+2)^2 \text{sech}^2\left(\frac{1}{2} t \sin (\theta )\right)\nonumber\\
    &+\left(t \tanh \left(\frac{1}{2} t \sin (\theta )\right)-2 \csc (\theta )\right)^2+4\bigg]\,.
\end{align}
\normalsize
Numerically, we can now compare the Krylov spread complexity $C_K(t)_\beta$ we derived in \eqref{krylov_result} to check whether it matches with $\langle\hat{L}\rangle_{\text{cl.}}$ in~\eqref{length} plus $\langle \hat{L}\rangle_{q\text{-corr.}}$ in~\eqref{eq:q-correction_length} for values of $q$ close to $q=1$. This comparison is visualized in Fig.~\ref{fig:correction}. We find excellent agreement around $q=1$, highlighting that the derived complexity indeed matches the quantum length. As expected, the agreement gets worse for lower $q$; this is because here, higher order corrections to the length that are not known analytically become relevant.

Furthermore, the functional form of the first quantum correction to the length \eqref{eq:q-correction_length} suggests a particular interpretation: In the early time limit, the correction itself behaves quadratically in $t$, supplemented by an onset in the case of finite inverse temperature $\beta$, while in the late time limit it exhibits linear growth. The functional behavior of this first quantum correction to the bulk length coincides with the characteristic behavior of Krylov spread complexities. On the basis of this observation, we raise the question whether the correction individually has a dual manifestation as a particular notion of Krylov spread complexity. Reminiscent of the case of holographic entanglement entropy where the first quantum correction is given by entanglement between bulk quantum fields~\cite{Faulkner:2013ana}, we speculate that the first length quantum correction has a dual interpretation as Krylov spread complexity of bulk quantum fields.\\


\noindent \textbf{\emph{Summary.--}} 
We presented a precise duality between Krylov spread complexity in DSSYK and length in sine dilaton gravity. This is accomplished by leveraging the freshly discovered duality between DSSYK and sine dilaton gravity to recast the bulk length as a Krylov spread complexity. In particular, to establish such a duality at finite temperature for the first time, gravity demands to assign complexity to the Euclidean preparation of the Hartle-Hawking state. The derived duality holds on the quantum level, meaning for all DSSYK parameters $q\in(0,1)$. This is exemplified by a numerical comparison of the derived Krylov spread complexity with the first quantum correction to the length.\\

\noindent \textbf{\emph{Outlook.--}} Krylov complexity is expected to saturate at times exponential in the entropy. Ab inito, this seems to conflict with the linear growth that is observed in \cite{Rabinovici:2023yex} and this letter as the asymptotic behavior of the studied Krylov spread complexities. The resolution of this puzzle lies in taking into account non-perturbative corrections. From the gravitational perspective, the two-point function is expected to receive contributions from higher topologies in the gravitational path integral of sine dilaton gravity. In particular, the saturation of the two-point function—manifesting as a late-time plateau that halts the linear growth—is likely driven by non-perturbative effects in the topological expansion. This mirrors the behavior observed in JT gravity, where the dual matrix model \cite{Saad:2019lba} provided a framework to explain such non-perturbative late-time phenomena \cite{Iliesiu:2021ari}. Remarkable progress in this direction was very recently also reported in \cite{Nandy:2024zcd,Balasubramanian:2024lqk}. We conjecture that similar mechanisms should govern sine dilaton gravity, whose dual matrix model is expected to correspond to the ETH matrix model for \( q \)-deformed JT gravity \cite{Jafferis:2022wez}, as recently proposed in \cite{Blommaert:2025avl}. On the boundary side, any potential incarnation of Krylov complexity is thus expected to go beyond the double-scaled SYK model, which corresponds to the strict \( N \to \infty \) limit.

Moreover, it is important to state that for Krylov spread complexity, a proof of the presence of the switchback effect—i.e. scrambling time delay of the linear growth associated with the presence of shockwaves in the bulk~\cite{Stanford:2014jda}—would complete its understanding as a genuine holographic complexity. There are strong reasons to believe that this characteristic can be made manifest in the context of the presented model. We expect that insights from JT gravity, such as the observation of Shapiro time delay in the two-point function of a heavy bilocal operator via a boundary clock analysis \cite{Mertens:2022irh} and the connection of the crossed four-point function with shockwave scattering  \cite{Lam:2018pvp}, translate to sine dilaton gravity. These characteristics are very closely tied to what would be interpreted as the switchback effect from the boundary complexity viewpoint, justifying the conjecture of existence of the switchback effect in DSSYK Krylov complexity. Recently, important boundary insights connecting to this line of argument have been reported for the triple-scaled DSSYK model in \cite{Xu:2024gfm,Ambrosini:2024sre}. A finite $\beta$ adaptation of these methods, together with relaxation of the triple-scaling limit, could facilitate the proposed study of the switchback effect in the general case that we elaborate on in this letter.

Furthermore, the analysis of the quantum correction to the length 
leads us to two important questions to be investigated. Firstly, it is important to gain a geometric understanding of the first quantum correction to the length in sine dilaton gravity. This carries the potential to obtain a refined understanding of the complexity equals volume conjecture in the full quantum regime where quantum corrections to the volume have to be taken into account. Secondly, the previously stated speculation that the first  quantum correction to the length has a dual interpretation as Krylov spread complexity of bulk quantum fields deserves further investigation.

Finally, the most important question is whether the paradigm presented in this letter extends to higher dimensions. An interesting lesson seems to be that a crucial ingredient in the match is to account for the Euclidean segment of time evolution.\\

\begin{acknowledgments}
We would like to thank Thomas G. Mertens, Andreas Belaey and Thomas Tappeiner for inspiring discussions that helped to spark this project and Mario Flory, Aranya Bhattacharya and Emiliano Rizza for collaborations on related questions. MPH acknowledges welcoming hospitality of Jagiellonian University and discussions with Romuald Janik during the completion of this project. TS is supported by the Research Foundation - Flanders (FWO) doctoral fellowship 11I5425N. JP acknowledges financial support from the European Research Council (grant BHHQG-101040024).
Funded by the European Union. Views and opinions expressed are however those of the author(s) only
and do not necessarily reflect those of the European Union or the European Research Council. Neither
the European Union nor the granting authority can be held responsible for them.
\end{acknowledgments}

\section{Supplemental material}

\noindent \textbf{\emph{SM1: More details on sine dilaton gravity and the duality.--}} 
Let us provide further details about the duality between sine dilaton gravity, described by the action in~\eqref{b+b}, 
 and double-scaled SYK. The ADM energy of sine dilaton gravity is given by \cite{Blommaert:2024ydx}:  
\begin{equation}  
E_{\mathrm{ADM}} = H_{\mathrm{grav}} = -\frac{\cos(\theta)}{2\left| \log q \right|}\,.  
\label{energy}  
\end{equation}  
We now aim to perform a canonical transformation in phase space to express the gravitational Hamiltonian in more convenient variables, following an approach analogous to the one used for JT gravity in \cite{Harlow:2018tqv}.  
In eq. (4), we computed the length of the thermofield double (TFD) state in the black hole geometry (3), expressed in terms of the two-sided Lorentzian time \( t \), which is canonically conjugate to the ADM energy \eqref{energy}.  
To proceed with canonical quantization, we derive the canonical conjugate variable \( P \) to the TFD length \( L \) (given in (4)) by imposing that the symplectic measure on phase space adopts the canonical form in terms of \( L \) and \( P \):  
\begin{equation}  
\omega = \mathrm{d}t \wedge \mathrm{d}H_{\mathrm{grav}} = \frac{1}{2\left| \log q \right|}\mathrm{d} L \wedge \mathrm{d} P\,.  
\label{symple}  
\end{equation}  
This is achieved through 
\begin{equation}  
P=\mathrm{i} \log \left( \mathrm{i}\sin(\theta)\tanh(\sin(\theta)t/2)+\cos(\theta)\right).
\end{equation}
By inverting \( \theta \) and \( t \) in terms of \( L \) and \( P \), we express the gravitational Hamiltonian \( H_{\mathrm{grav}} \) in the new variables:  
\begin{equation}  
H_\text{grav}(L, P) = -\frac{\cos(P)}{2\left| \log q \right|} + \frac{1}{4\left| \log q \right|} e^{iP} e^{-L}.  
\label{grav}  
\end{equation}  
Upon quantization of this Hamiltonian, operator ordering ambiguities may arise. A particular ordering choice, replacing $ 
e^{\mathrm{i}\hat{P}}\rightarrow e^{\alpha(q) \hat{L}} e^{\mathrm{i}\hat{P}}e^{-\alpha(q)\hat{L}}
$ for $\alpha(q)=\frac{1}{4 \left|\log q\right|} \log \left(\frac{2 \left|\log q\right|}{1-q^2}\right)$ (we thank the authors of \cite{Bossi:2024ffa} for a discussion about this point), yields the following quantum Hamiltonian  
\small\begin{equation}  
\hat{H}_\text{grav}(\hat{L}, \hat{P}) = \frac{1}{\sqrt{2 \left|\log q\right|(1 - q^2)}}  \left[-\cos(\hat{P}) + \frac{1}{2} e^{i \hat{P}} e^{-\hat{L}} \right]  
\label{grav3}  
\end{equation} 
\normalsize
with spectrum
\begin{equation}\label{eigenvalue}
E(\theta) = - \cos(\theta)/\sqrt{2 \abs{\log q} (1 - q^2)}\,.
\end{equation}
This Hamiltonian exactly corresponds to the DSSYK transfer matrix \( \hat{T} \), in accordance with the holographic dictionary (6). The mapping  
\begin{equation} 
a_{\mathrm{norm}}, a^{\dagger}_{\mathrm{norm}} = - e^{\pm i \hat{P}}  
\end{equation} 
relates to the normalized raising and lowering operators in the chord basis \cite{Lin:2022rbf, Berkooz:2024lgq}.  
In the gravitational framework, the symplectic structure \eqref{symple} induces the standard commutation relation  
$[\hat{L}, \hat{P}] = 2i \left| \log q \right|$, implying that \( e^{i \hat{P}} \) acts as a shift operator by \( 2 \left| \log q \right| \). Consequently, the Schrödinger equation associated with \eqref{grav3} takes the form of a difference equation in the \( L \)-basis:  
\begin{equation}  
2 E(\theta) \psi_\theta(L) = \psi_\theta(L + 2\left|\log q \right|) + (1 - e^{-L}) \psi_\theta (L - 2\left| \log q \right|).  
\label{scho2}  
\end{equation}  
This equation can be solved for a general continuous variable \( L \) and after discretization \cite{Blommaert:2024whf} it reduces to the DSSYK wavefunctions in the $L$-basis given by q-Hermite polynomials
\begin{equation}\label{wave}
\psi_\theta(L)=\langle \theta | L \rangle=H_{n}\left(\cos(\theta)|q^2\right), 
\end{equation}
with right eigenvectors given instead by $\langle L | \theta \rangle=H_{n}\left(\cos(\theta)|q^2\right)/\left(q^2;q^2\right)_n $ \cite{Blommaert:2023opb} \textcolor{black}{where the q-Pochhammer symbol is defined as $(x;q^2)_n=\prod_{k=0}^{n-1}(1-x q^{2k})$ and the q-Hermite polynomials are the polynomials that solve the recursion relation $2x H_n(x\vert q^2) = H_{n+1} (x\vert q^2) + (1-q^{2n}) H_{n-1} (x\vert q^2)$ subject to the initial conditions $H_{-1}(x\vert q^2) = 0$ and $ H_0(x\vert q^2) = 1$}. This shows that the global slice Wheeler-DeWitt wavefunctions of sine dilaton gravity are the solutions of the Schroedinger equation associated with \eqref{grav3}. Furthermore, the orthogonality of these wavefunctions yields the DSSYK spectral density via:  
\begin{equation} 
\begin{split}
\langle \theta | \theta' \rangle &= \sum_{n=0}^{\infty} \frac{1}{(q^2; q^2)_n} H_{n}(\cos(\theta) | q^2) H_{n}(\cos(\theta') | q^2)  \\
&= \frac{\delta(\theta - \theta')}{(e^{\pm 2i \theta}; q^2)_\infty}.
\end{split}
\end{equation}\\

\noindent \textbf{\emph{SM2: Analytic expression for the two-point function of a massive probe in sine dilaton gravity.--}}
By inserting two completeness relations 
$\int_0^{\pi}\rho(\theta)\ket{\theta}\bra{\theta}=\mathbb{1}$ into~\eqref{2exact}, 
we obtain
\begin{align}\label{two}
   \langle e^{-\Delta \hat{L}} \rangle &=Z_{\beta}^{-1}\int_0^\pi \mathrm{d}\theta_1\,\rho(\theta_1)\int_0^\pi \mathrm{d}\theta_2\,\rho(\theta_2)\exp\left(-\tau E(\theta_1)\right)\nonumber\\
&\exp\left(-(\beta-\tau)E(\theta_2)\right) \bra{\theta_1} e^{-\Delta \hat{L}} \ket{\theta_2}\,,
\end{align}
where one has \cite{Berkooz:2018jqr,Blommaert:2023opb} 
\begin{align}
    &\bra{\theta_1} e^{-\Delta \hat{L}} \ket{\theta_2}=\frac{(q^{4\Delta};q^2)_\infty}{(q^{2\Delta}e^{\pm2i\theta_1\pm2i\theta_2};q^2)_\infty}\nonumber\\
    &=\sum_{n=0}^{\infty} \frac{q^{2n\Delta}}{(q^2; q^2)_n} H_n \left( \cos \theta_1 | q^2 \right) H_n \left( \cos \theta_2 | q^2 \right)\,,
\label{eq:matrixelement_sum}
\end{align}
the energy eigenvalues are given by \eqref{eigenvalue} and $\rho(\theta)=(q^2, e^{\pm 2\i\theta};q^2)_\infty
$ is the spectral density associated with the wavefunctions \eqref{wave}. \textcolor{black}{Here, we used common shorthand notation to abbreviate the product of q-Pochhammer symbols $(q^2, e^{\pm 2\i\theta};q^2)_\infty=(q^2;q^2)_\infty(e^{+ 2\i\theta};q^2)_\infty(e^{- 2\i\theta};q^2)_\infty
$ and notice again the appearance of the q-Hermite polynomials. Both q-Pochhammer symbols and q-Hermite polynomials are definied in the previous section of the supplemental material.}\\

\noindent \textbf{\emph{SM3: Complexity equals volume:  operator statement.--}}
Let us highlight that Krylov spread complexity equals wormhole length in sine dilaton gravity dual to finite temperature DSSYK can also be made as an operator statement. Krylov spread complexity as introduced in~\eqref{eq:Krylov_complexity} 
can be rewritten as the expectation value 
\begin{align}
    C_K(t)=\langle \psi(t)|\hat{K}|\psi(t)\rangle\,,\quad \hat{K}=\sum_{n=0}^{\infty} n\, \vert K_n\rangle\langle K_n\vert
\end{align}
of the Krylov operator $\hat{K}$ in the state $\ket{\psi(t)}$ that, in the setup of this letter, arises via transfer matrix evolution from the zero chord state as reference state. For this specific evolution, the Krylov basis $|K_n\rangle$ coincides with the chord basis $|n\rangle$ in DSSYK \cite{Rabinovici:2023yex} for times smaller than exponential in the entropy which is the natural restriction of the presented argument. Through the holographic dictionary $L=2\left| \log q\right| n$ for sine dilaton gravity, established in \cite{Blommaert:2024whf} and discussed in the setup, it is also known that the chord number operator $\hat{n}=\sum_{n=0}^{\infty} n \vert n\rangle\langle n\vert$ is the length operator $\hat{L}$ in this specific theory of gravity. Combining these two statements implies that in sine dilaton gravity dual to finite temperature DSSYK, the Krylov operator coincides with the length operator 
\begin{align}
\label{app:op_eq}
    2 \left| \log q\right| \hat{K} = \hat{L}\,.
\end{align}
This is an operator manifestation of complexity equals volume in the setup of this letter. Compared to previous statements in \cite{Lin:2022rbf,Rabinovici:2023yex} the novelty of this statement lies in the fact that we precisely state the interpretation of $\hat{L}$ dual to the DSSYK Krylov operator: It is the length operator in sine dilaton gravity. By looking at the expectation value of \eqref{app:op_eq} in the state $\ket{\psi(t)}$ that arises via transfer matrix evolution from the zero chord state, we find
\begin{align}
    2\left| \log q\right| C_K(t)=\langle \psi(t)|\hat{L}|\psi(t)\rangle
\end{align}
which precisely agrees with the result we have derived by explicit calculation in~\eqref{krylov_result}.\\

\noindent \textbf{\emph{SM4: SYK interpretation of the chord Krylov problem.--}} In~\eqref{krylov_result} we identified wormhole length in sine dilaton gravity with DSSYK Krylov spread complexity, computed in the auxilary quantum system of the chord Hilbert space governed by transfer matrix evolution. The assoiciated Krylov problem to the complexity we have derived is 
\begin{align}
\label{app_eq:Krylovproblem_chord}
    \ket{\psi(t)}_\beta=Z_\beta^{-1/2}e^{-iT(t - i \beta/2)} | 0 \rangle
\end{align}
where the reference state is the zero chord state $\ket{0}$ and we consider mixed Lorentzian plus Euclidean evolution with the transfer matrix $T$. Let us comment on how to translate this story from the chord construction back to the double scaled SYK model, governed by the Hamiltonian $H_{SYK}$, as introduced in (1), in the double scaling limit.

As the authors of \cite{Rabinovici:2023yex} have pointed out, the fact that Krylov spread complexity is determined by the moments of the evolution Hamiltonian in the reference state \cite{Balasubramanian:2022tpr} (or equivalently the survival amplitude $\braket{\psi(0)}{\psi(t)}$), makes it possible to translate the chord Krylov problem \eqref{app_eq:Krylovproblem_chord} back into a SYK statement. The foundation of this argument is that, by construction of the chord method, coupling averaged moments of the SYK Hamiltonian in the double-scaling limit are calculated via moments of the transfer matrix in the zero chord state \cite{Berkooz:2018jqr} 
\begin{align}
    \left\langle \Tr[(H_{SYK})^{2k}] \right\rangle_J = \bra{0} T^{2k} \ket{0}\,.
\end{align}
By rewriting the trace, one can now identify the state $\ket{\Omega}=\mathcal{N}^{-1/2}\sum_E\ket{E}$, an equal superposition of energy eigenstates $\ket{E}$ of $H_{SYK}$ where $\mathcal{N}$ is a normalization constant, as the reference state on the level of double-scaled SYK. The associated Krylov problem is equivalent to the  chord Krylov problem after averaging over the coupling because 
\begin{align}
    \left\langle\bra{\Omega} (H_{SYK})^{2k} \ket{\Omega}\right\rangle_J = \bra{0} T^{2k} \ket{0}\,.
\end{align}
Applying this argument of \cite{Rabinovici:2023yex} to the finite temperature construction in this letter, we identify the Krylov problem on the level of SYK that is equivalent to \eqref{app_eq:Krylovproblem_chord} as
\begin{align}
\label{app_eq:Krylovproblem_SYK}
    \ket{\Psi(t)}_\beta=Z_\beta^{-1/2}e^{-iH_{SYK}(t - i \beta/2)} | \Omega \rangle\,,
\end{align}
where $\ket{\Omega}=\mathcal{N}^{-1/2}\sum_E\ket{E}$ is the reference state and we consider mixed Lorenztian plus euclidean evolution with the double-scaled SYK Hamiltonian. Let us point out that the argument of \cite{Rabinovici:2023yex} is also applicable in the finite temperature case only because, as stated in the main text, we choose to assign complexity to both Lorentzian evolution as well as Euclidean state preparation and do not include the Euclidean state preparation in the reference state.\\

\noindent \textbf{\emph{SM5: Krylov spread complexity: Numerical evaluation.--}} For transparency, let us lay out briefly how we obtain data points in our numerical evaluation of Krylov spread complexity at finite $\beta$ and generic $q\in(0,1)$. Numerical evaluation is facilitated by introduction of two cutoffs $n_\text{cut}$ and $k_\text{cut}$ (chosen sufficiently large) in the relevant infinite sums. We numerically evaluate 
\begin{align}
    C_K(t)_\beta=Z_\beta^{-1}\sum _{n=1}^{n_\text{cut}} \frac{n }{(q^2;q^2)_n}| \phi_n(t-i\beta/2)|^2
\end{align}
where
\begin{align}
    &\phi_0(t)=\sum _{k=-k_\text{cut}}^{k_\text{cut}} q^{2\binom{k}{2}} J_{2 k}\left(\frac{ t/\sqrt{1-q^2}}{\sqrt{-2\log (q)} }\right)\,,\\
    &\phi_n(t)=(-i)^n \sum _{k=0}^{k_\text{cut}} \bigg[\frac{q^{2\binom{k}{2}} \left(1-q^{2(2 k+n)}\right) (q^2;q^2)_{k+n} }{\left(1-q^{2(k+n)}\right) (q^2;q^2)_k}\nonumber\\
    &\quad\quad\quad\quad\quad\quad\quad\quad\quad\times J_{2 k+n}\left(\frac{t/\sqrt{1-q^2}}{\sqrt{-2\log (q)} }\right)\bigg]\,.
\end{align}
We owe these analytic expressions for the wave functions  $\phi_n(t)$, which are key for the numerical implementation, to the detailed analysis in \cite{Xu:2024gfm}. The infinite sum in $k$ converges sufficiently fast (for thorough analysis see \cite{Xu:2024gfm}) such that cutting off the sum at $k_\text{cut}$ large enough is justified. We have chosen $k_\text{cut}=50$ in our implementation. A sufficiently large cutoff $n_\text{cut}$ can now be identified by requiring the norm of the state 
\begin{align}
    1\overset{!}{=}Z_\beta^{-1}\sum _{n=0}^{n_\text{cut}} \frac{1}{(q^2;q^2)_n}| \phi_n(t-i\beta/2)|^2
\end{align}
to agree with one within a desired accuracy. Here, $Z_\beta$ also depends on \textcolor{black}{$k_\text{cut}$ in this implementation because
\begin{align}
    Z_\beta=\phi_0(-i\beta)\,.
\end{align}}

\bibliographystyle{bibstyl}
\bibliography{V4k-complexity.bib}

\begin{thebibliography}{10}
\bibitem{footnote3}
Without loss of generality, we set the DSSYK coupling $J$ that appears as a multiplicative factor in $E(\theta)$ to $J=1$ for the rest of this letter. Hence, all times are understood to be measured in units of $J$.
\bibitem{footnote4}
We point out that the $q\to1$ limit in our framework is different from the triple scaling limit $(q\to1,\theta\ll1)$, since we are keeping $\theta$ fixed. We regard it as a semiclassical limit because, after rescaling the dilaton in the sine dilaton action \eqref{b+b} as $2 \abs{\log q} \Phi\rightarrow \Phi$ which is our convention in this letter, $\abs{\log q}$ stands on the same footing as $G_{\mathrm{N}}$.
\end{thebibliography}

\end{document}